# Comparative study of *in situ* and *ex situ* MgB$_2$ films deposited by pulsed laser deposition

Y. Zhao, M. Ionescu, J. Horvat, and S. X. Dou
Institute for Superconducting and Electronic Materials, University of Wollongong, NSW 2522, Australia
E-mail: yz70@uow.edu.au

**Abstract**: Two types of MgB$_2$ films were prepared by pulsed laser deposition (PLD) with *in situ* and *ex situ* annealing processes respectively. Significant differences in properties between the two types of films were found. The *ex situ* MgB$_2$ film has a T$_c$ of 38.1K, while the *in situ* film has a depressed T$_c$ of 34.5K. The resistivity at 40K for the *in situ* film is larger than that of the *ex situ* film by a factor of 6. The residual resistivity ratios (RRR) are 1.1 and 2.1 for the *in situ* and *ex situ* films respectively. The J$_c$-H curves of the *in situ* film show a much weaker field dependence than those of the *ex situ* film, attributable to stronger flux pinning in the *in situ* film. The small-grain feature and high oxygen level may be critical for the significant improvement of J$_c$ in the *in situ* annealed MgB$_2$ film.

Shortly after the discovery of superconductivity in MgB$_2$, thin films with bulk-like T$_c$ were achieved by *ex situ* annealing of a boron or Mg-B precursor film in Mg vapor.[1,2] Significantly higher J$_c$ at zero field was obtained in the *ex situ* thin film than in bulk, however the J$_c$ dropped sharply with increasing field as was observed in Ref. 1. Kang and co-workers optimized the *ex situ* annealing conditions by fine-tuning the annealing temperature and time, and a very high J$_c$ of ~10$^7$J/cm$^2$ at 5K, 0T, with a weak field dependence was obtained.[3] Eom and coworkers found that an high oxygen level in the MgB$_2$ film prepared by a two-step procedure significantly improved the J$_c$ in high fields.[2] The technical drawback for applications of the two-step *ex situ* procedures is the inability to produce multi-layer structures. Zeng et al. developed an *in situ* HPCVD method to grow high quality MgB$_2$ film epitaxially.[4] The J$_c$ value of their MgB$_2$ film reached 3.5x10$^7$A/cm$^2$ at 4.2K, 0T (transport measurement), but it dropped rather quickly to 10$^5$A/cm$^2$ as the field increased to 4T, which indicates weak pinning due to a low level of impurity and disorder in their films.[4] Other MgB$_2$ films, as-grown and *in situ* annealed, have more or less suppressed T$_c$,[5-8] but the J$_c$ properties are quite good in high fields at lower temperature.[5] It seems that the differences in superconducting properties between different types of films strongly depend on the microstructure and preparation processes. However, the detailed mechanisms are still not clear.

In this letter, we report a comparative study on *in situ* and *ex situ* annealed MgB$_2$ films prepared by pulsed laser deposition. Significant enhancement of the pinning force as well as a weak field dependence of J$_c$ is obtained in the *in situ* annealed film. The differences in both microstructure and superconductivity between the *in situ* and *ex situ* annealed MgB$_2$ films are discussed.

In the preparation of the *in situ* annealed film, the precursor film was deposited on an Al$_2$O$_3$-R substrate from a stoichiometric MgB$_2$ target (84% density). The pulsed laser (248nm, 300mJ/pulse) was focused to an elliptical spot with dimensions of 7mm x 1.5mm on the target. The deposition time for MgB$_2$ is 5min with a laser repetition of 10Hz. During the deposition, the substrate was kept at 250℃ and in Ar at 120mTorr. A ~800nm thick Mg cap layer was then deposited onto the precursor film. The film was heated *in situ* to 685 ℃ in 12 min and kept at this temperature for 1min in a 1 atm Ar atmosphere.[5] For the *ex situ* annealed MgB$_2$ film, a boron precursor film was deposited from a boron target (~40% density) in a 10$^{-7}$-10$^{-6}$ Torr vacuum. The substrate was also kept at 250℃ during the deposition. The precursor film was then wrapped in Ta foil and sealed in a stainless steel tube in an Ar atmosphere, together with Mg pellets. The tube was put into a 900℃ furnace and kept for 30min. The *ex situ* annealing conditions follow Kang et al.'s optimized procedure.[3] We prepared each type of MgB$_2$ film several times under the same conditions, and good reproducibility in T$_c$ was found in both types of films.

The transport measurement was carried out on a PPMS-9T magnetometer system (Quantum Design), using a standard 4-probe method. The zero-field-cooled (ZFC) magnetization vs. temperature (M-T) curves and magnetization hysteresis loops of the films were measured on an MPMS-5T magnetometer. In each measurement the applied field was perpendicular to the film plane. J$_c$(A/cm$^2$) was calculated from the height of magnetization loops, $\Delta$M(emu/cm$^3$), using the standard equation based on the Bean model: J$_c$=20$\Delta$M/[a(1-a/3b)]. Due to the similarity in dimensions of our films (0.55cm x 0.25cm x 350nm), the J$_c$ obtained can be used for comparative purposes. The irreversibility field (H$_{irr}$) and H$_{c2}$ at each temperature were selected as the point of 10% and 90% of the normal state resistivity in the resistivity-T curves, respectively.

The surface topography and the thickness of the films were determined by both atomic force microscopy (AFM) and scanning electron microscopy (SEM). The cross-sectional SEM images of the two films are shown in Figure 1. We take 350nm as the thickness of the *in situ* annealed film in the calculations of J$_c$ and resistivity, but this value is actually a lower estimate of the effective thickness of the *in situ* film due to the existence of big islands on the film (Figure 1(a)). The thickness of the *ex situ* annealed film is also about 350nm, which is more accurate because of the comparatively smooth and homogeneous surface of the film (Figure 1(b)).

Figure 2 shows the resistivity vs. temperature curves of the two films. The $T_c$ is 38.1K ($\Delta T=1K$) and 34.5K ($\Delta T=3K$) for the *ex situ* and *in situ* annealed films respectively. The *ex situ* annealed film has a higher $T_c$ and a narrower transition width in the transport measurements. However, we notice that the bulk diamagnetism transition in the M-T curve for the *in situ* film ($\Delta T=5K$) is actually sharper than for the *ex situ* annealed film ($\Delta T=9K$), which indicates a more homogeneous superconducting phase in the *in situ* film.

Also shown in Figure 2 is a significant difference between the resistivity of the two types of films. The resistivity of the *in situ* annealed film is 104 µΩcm at 40K, whereas the resistivity of the *ex situ* film is 17µΩcm. The residual resistivity ratios are 1.1 and 2.1 for the *in situ* and the *ex situ* annealed films respectively. The *ex situ* annealed $MgB_2$ film showed more metallic properties than the *in situ* one in terms of electrical transport in the normal state.

Figure 3 shows the temperature dependence of $H_{irr}$ and $H_{c2}$ for the $MgB_2$ films. Data for $MgB_2$ films from the literature are shown for comparison.[2,9] The slope of the $H_{irr}$ ($H_{c2}$)-T curve of the *in situ* annealed film is significantly higher than for the *ex situ* film. The $H_{irr}$ values and the slope of $H_{irr}$-T curve of our *in situ* annealed film are higher than the oxygen alloyed *ex situ* film reported by Eom et al,[2] which indicates a further enhancement of pinning force in our *in situ* annealed film. Figure 3(b) shows a comparison of $H_{c2}$ between the *in situ* and *ex situ* films. There is a good agreement with the literature in that films with higher resistivity have larger $H_{c2}$-T slopes.[9] The $H_{c2}$ value for our *in situ* film is higher than that of the oriented high resistivity film studied in Ref. 9, with the applied field perpendicular to the film plane.

The field dependence of $J_c$ of the two films is shown in Figure 4. The critical current is over $5 \times 10^6 A/cm^2$ in zero field from 5 to 20 K for both *in situ* and *ex situ* films. The *in situ* annealed $MgB_2$ film shows a much weaker field dependence of $J_c$ than the *ex situ* film. The $J_c$ at 5K and 5 T is about $10^6 A/cm^2$, whereas the $J_c$ for the *ex situ* film drops to less than $10^4 A/cm^2$ in the same field and temperature. The $J_c$ behavior of the *ex situ* annealed $MgB_2$ film is very similar to that of clean $MgB_2$ bulks, which indicates a lack of pinning in high fields.

A clear difference in the surface features of the two films is revealed by AFM images, as shown in Figure 5. The *ex situ* annealed film showed a typical crystallized surface with random grain orientation. The individual blocks, which are probably crystallites, are about 400 nm in size. This grain size is consistent with the cross-section observation in Figure 1(b). In contrast, the detailed scan of the *in situ* annealed film, shown in Figure 5(a), reveals small rounded cells of 30-100 nm in diameter. Considering the lack of crystalline surface characteristics, it is possible that inside these cells there are even smaller grains. The clear microstructural differences between the two films may be of decisive importance for the significant difference between the field dependence of their $J_c$. In the *in situ* films, the small grains with their grain boundaries, usually about 3nm wide, are of desirable dimensions for efficient pinning.[10] Whereas in the *ex situ* film, the grain boundary pinning is less pronounced for the relatively larger grains.

Impurities may also play an important role in the increase in $H_{c2}$ and the enhancement of flux pinning in the *in situ* film. As shown in the EDS result (Figure 6), the oxygen level in the *in situ* annealed film is obviously higher than in the *ex situ* annealed film. Due to the high reactivity of Mg with oxygen, a significant amount of oxygen may be introduced into the *in situ* film during the $MgB_2$ deposition process. For the *ex situ* film, the boron precursor is much less sensitive to oxygen. As discussed in the literature,[2,9] the oxygen in $MgB_2$ bulk and thin film could substitute for boron or exist as MgO precipitates. In both circumstance the intra-band scattering is enhanced, and effectively increases the slope of the $H_{c2}$-T curve for the $MgB_2$ thin film.[9] Despite the high oxygen level, the $T_c$ of our in situ film was not dramatically suppressed, which is consistent with the theoretical prediction that the inter-band scattering rate between σ and π band would be small and the $T_c$ less influenced by intra-band scattering in $MgB_2$.[11] Moreover, nano-sized MgO particles, which are usually observed in oxygen-rich $MgB_2$ films,[2] could provide effective pinning centers, and lead to a further enhancement of critical current density in high fields.

The authors thank E. W. Collings for help in providing the stoichiometric $MgB_2$ target. This work is supported by Australian Research Council (ARC) under a Linkage Project (LP0219629) cooperating with Alphatech International and The Hyper Tech Research Inc.

[1] W N Kang, H J Kim, E M Choi, C U Jung and S I Lee, Science **292**, 1521 (2001).

[2] C.B. Eom, M.K. Lee, J.H. Choi, L.J. Belenky, X. Song, L.D. Cooley, M.T. Naus, S. Patnaik, J. Jiang, M. Rikel, A. Polyanskii, A. Gurevich, X.Y. Cai, S.D. Bu, S.E. Babcock, E.E. Hellstrom, D.C. Larbalestier, N. Rogado, K.A. Regan, M.A. Hayward, T. He, J.S. Slusky, K. Inumaru, M.K. Hass and R.J. Cava, Nature **411**, 558 (2001).

[3] W.N. Kang, E.-M. Choi, H.-J. Kim, H.-J. Kim and S.-I. Lee, cond-mat/0209226

[4] X.H. Zeng, A.V. Pogrebnyakov, M.H. Zhu, J.E. Jones, X.X. Xi, S.Y. Yu, E. Wertz, Q. Li, J.M. Redwing, J. Lettiery, V. Vaithyanathan, D.G. Schlom, Z.K. Liu, O. Trithaveesak and J. Schubert, Appl. Phys. Lett. **82**, 2097 (2003).

[5] Y. Zhao, M. Ionescu, A. V. Pan, S. X. Dou and E. W. Collings, Supercond. Sci. Technol. **16**, 1487 (2003), in press.

[6] D.H.A. Blank, H. Hilgenkamp, A. Brinkman, D. Mijatovic, G. Rijnders and H. Rogalla, Appl. Phys. Lett. **79**, 394 (2001).

[7] H.M. Christen, H.Y. Zhai, C. Cantoni, M. Paranthaman, B.C. Sales, C. Rouleau, D.P. Norton, D.K. Christen and D.H. Lowndes, Physica C **353**, 157 (2001).

[8] K. Ueda and M Naito, Appl. Phys. Lett. **79**, 2046 (2001).

[9] A. Gurevich, S. Patnaik, V. Braccini, K.H. Kim, C. Mielke, X. Song, L.D. Cooley, S.D. Bu, D.M. Kim, J.H. Choi, L.J. Belenky, J. Giencke, M.K. Lee, W. Tian, X.Q. Pan, A. Siri, E.E. Hellstrom, C.B. Eom and D.C. Larbalestier, cond-mat/0305474.

[10] Y. Bugoslavsky, L. Cowey, T.J. Tate, G.K. Perkins, J. Moore, Z. Lockman, A. Berenov, J. LMacManus-Driscoll, A.D. Caplin, L.F. Cohen, H.Y. Zhai, H.M. Christen, M.P. Paranthaman, D.H. Lowndes, M.H. Jo and M.G. Blamire, Supercond. Sci. Technol. **15**, 1392 (2002).

[11] I.I. Marzin and V.P. Antropov, cond-mat/0212263.

**Figure captions:**

Figure 1. SEM cross-section images of the two types of films. a: *in situ* annealed $MgB_2$ film; b: *ex situ* annealed $MgB_2$ film

Figure 2. Temperature dependence of the resistivity of the two types of $MgB_2$ films from 5K to 300K in zero field.

Figure 3. Irreversibility lines (IL) and upper critical field versus temperature curves for the *in situ* and *ex situ* annealed films. a: Irreversibility lines for the *in situ* and *ex situ* films. The data from the literature (Ref. 2) are put in the figure for comparison. b: Upper critical fields versus temperature for the two films. The data from Ref. 9 are shown in this figure.

Figure 4. Critical current of the two types of $MgB_2$ films calculated from M-H loops. Solid symbols: *ex situ* annealed film; open symbols: *in situ* annealed film.

Figure 5. AFM images (illuminated 3D mode) of the *in situ* and *ex situ* annealed MgB2 films. (a) *in situ* annealed film in 500x500 $nm^2$ range, (b) *ex situ* annealed film in 2x2 $\mu m^2$ range.

Figure 6. EDS peaks for the *in situ* and *ex situ* films. In order to avoid interaction with the $Al_2O_3$ substrate, a low electron beam energy of 5kV was used. (a) *in situ* film, (b) *ex situ* film.

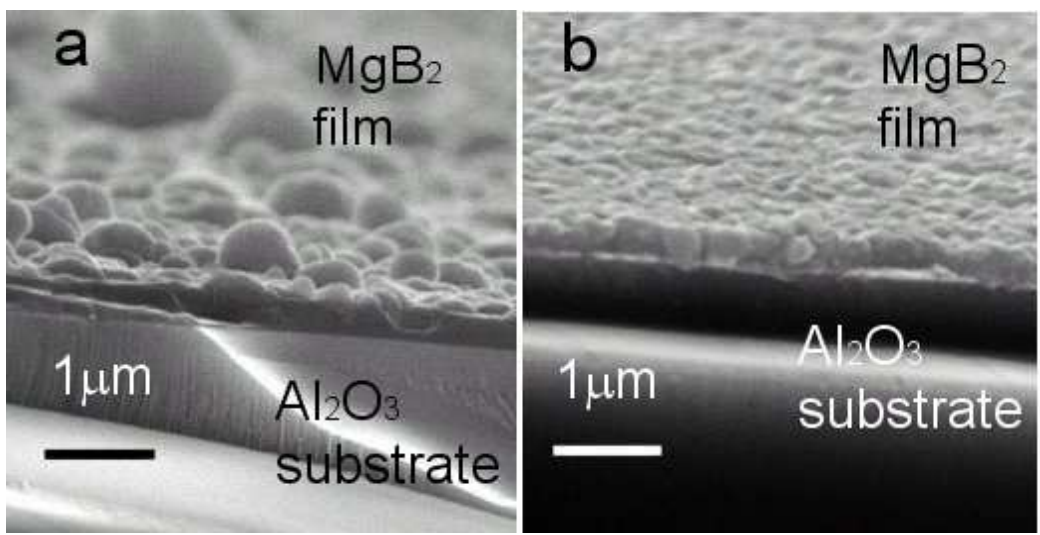

Figure 1

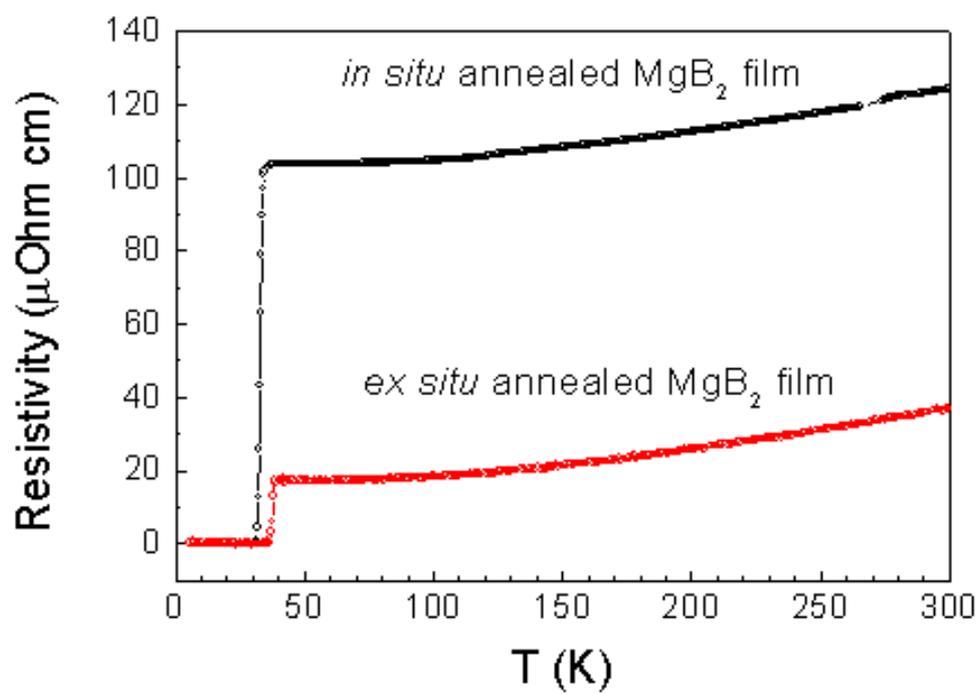

Figure 2

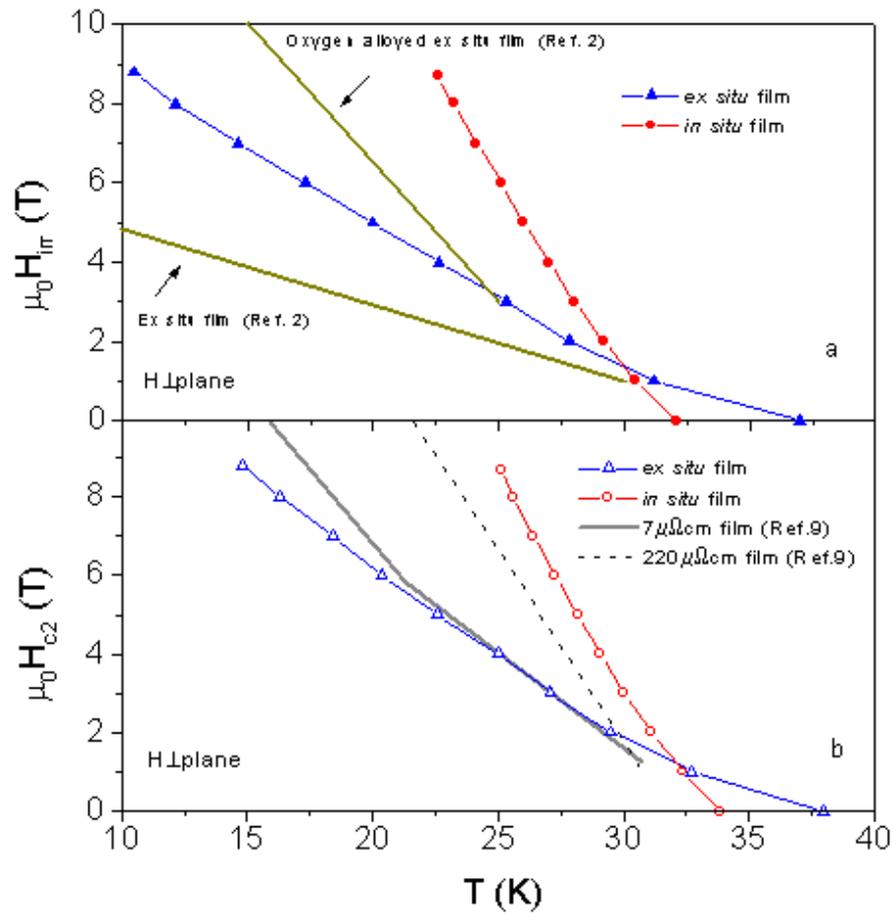

Figure 3

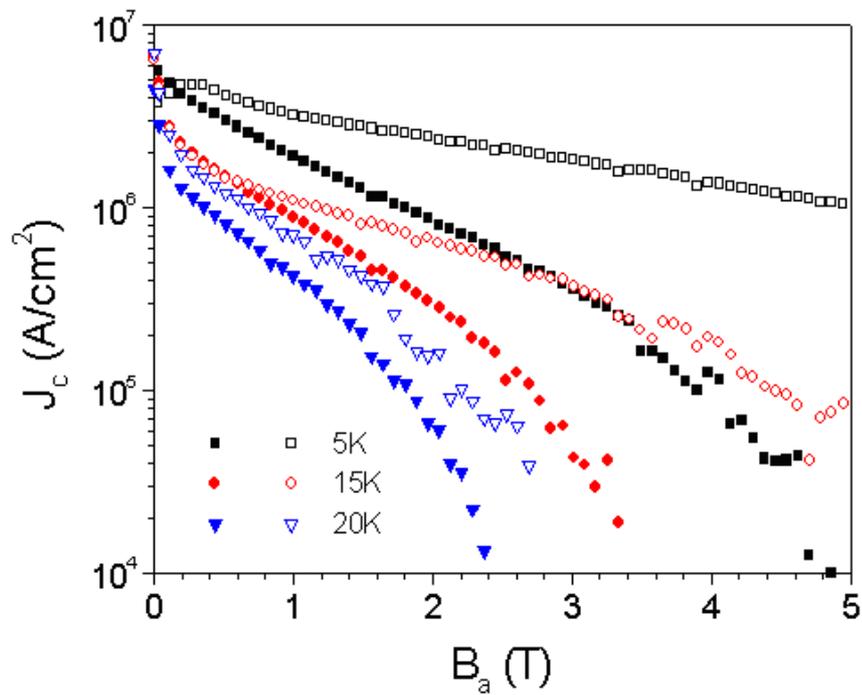

Figure 4

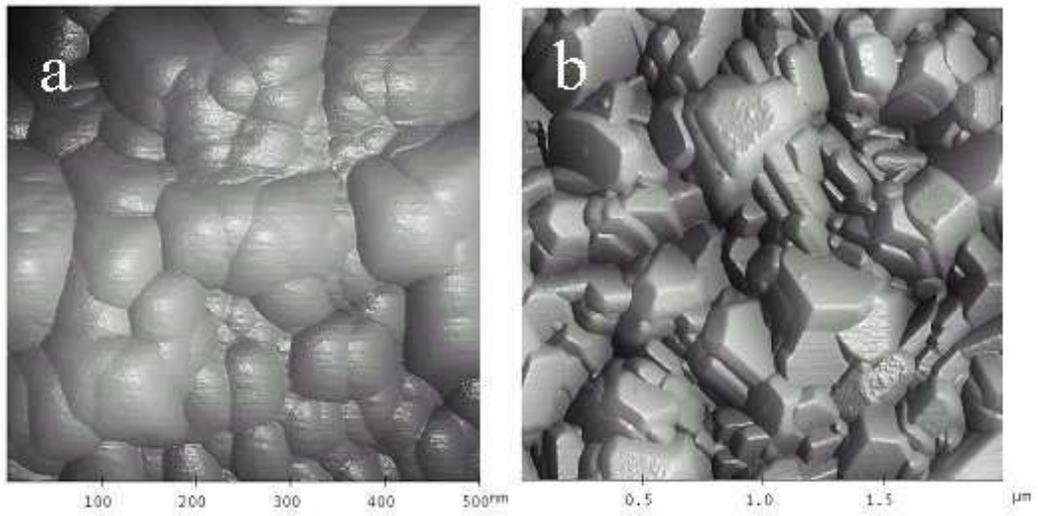

Figure 5

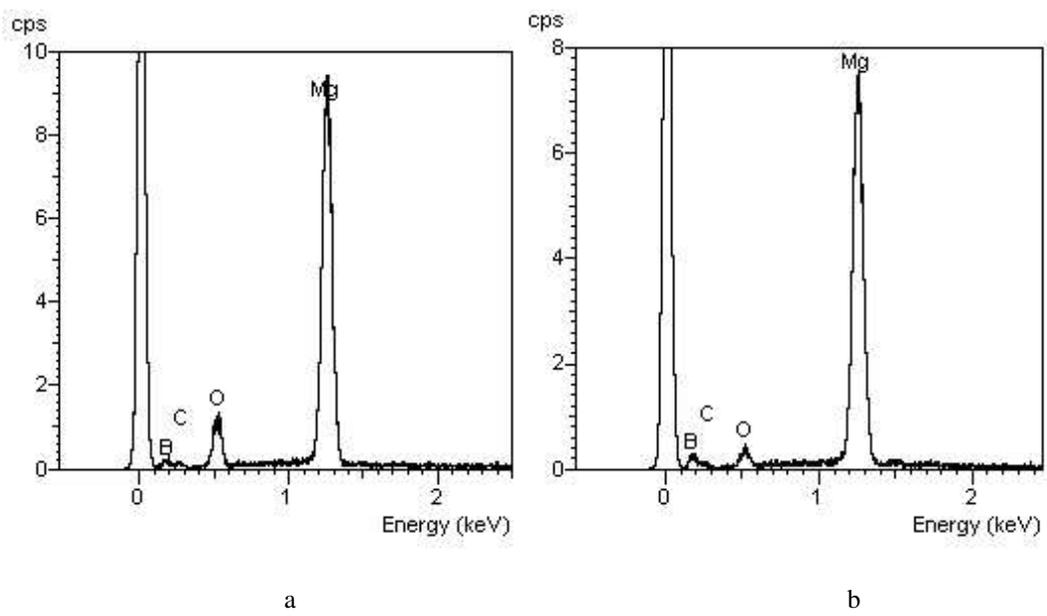

a					b

Figure 6